# The Orion HII Region and the Orion Bar in the Mid-Infrared

F. Salgado[1], O. Berné[2,3], J. D. Adams[4], T. L. Herter[4], L. D. Keller [5] and A. G. G. M. Tielens[1]

## ABSTRACT

We present mid-infrared photometry of the Orion Bar obtained with FORCAST aboard SOFIA at 6.4, 6.6, 7.7, 19.7, 31.5 and 37.1 $\mu$m. By complementing this observations with archival FORCAST and *Herschel*/PACS images we are able to construct a complete infrared spectral energy distribution of the Huygens region in the Orion nebula By comparing the infrared images with gas tracers, we find that PACS maps trace the molecular cloud, while the FORCAST data trace the photodissociation region (PDR) and HII region. Analysis of the energetics of the region reveal that the PDR extends for 0.28 pc along the line-of-sight and that the Bar is inclined at an angle of 4°. The infrared and submillimeter images reveal that the Orion Bar represents a swept up shell with a thickness of 0.1 pc. The mass of the shell implies a shock velocity of $\simeq 3$ km/s and an age of $\simeq 10^5$ yr for the HII region. Our analysis shows that the UV and infrared dust opacities in the HII region and the PDR are a factor 5 to 10 lower than in the diffuse interstellar medium. In the ionized gas, Ly$\alpha$ photons are a major source of dust heating at distances larger than $\simeq 0.06$ pc from $\theta^1$ Ori C. Dust temperatures can be explained if the size of the grains is between 0.1 to 1 $\mu$m. We derive the photo-electric heating efficiency of the atomic gas in the Orion Bar. The results are in good qualitative agreement with models and The quantitative differences indicate a decreased PAH abundance in this region.

*Subject headings:*

## 1. Introduction

Dust grains are an important constituent of the interstellar medium (ISM) of galaxies. From young stellar objects (YSOs) to active galactic nuclei (AGN), most of the objects in the Universe show the presence of dust either in emission or absorption. Moreover, dust plays a key role in star and planet formation, molecule formation due to surface chemical reactions and heating of the gas due to the photoelectric effect. In addition, dust emission provides a convenient tracer of star formation in galaxies (e.g. Calzetti et al. 2007, 2010).

[1]Leiden Observatory, University of Leiden, P. O. Box 9513, 2300 RA Leiden, Netherlands

[2]Université de Toulouse; UPS-OMP; IRAP; Toulouse, France

[3]CNRS; IRAP; 9 Av. colonel Roche, BP 44346, F-31028 Toulouse cedex 4, France

[4]Astronomy Department, 202 Space Sciences Building, Cornell University, Ithaca, NY 14853-6801, USA

[5]Department of Physics and Astronomy, Ithaca College, Ithaca, NY 14850, USA

There is a consensus that dust grains are mainly composed of two chemically different type of species: silicates and carbonaceous materials (Zubko et al. 2004; Draine & Li 2007). The former being the most abundant in mass for large grains, and the latter being the dominant species for smaller grains. At the lower end of the size distribution of grains, large molecules known as poly-cyclic aromatic hydrocarbons (PAH), formed by carbon rings, produce emission in characteristic infrared bands from 3 to 20 $\mu$m (Tielens 2008). While the general properties of the dust grains in the diffuse ISM have been well studied, dust grain properties are expected to change with the environment. In particular, measurements of the extinction curve towards different lines of sight in the Galaxy (Cardelli et al. 1989) reveal changes that are consistent with changes in the size distribution of dust grains. This finding is further supported by studies such as Stepnik et al. (2003) have also demonstrated variations in the dust size distribution and in the abundance measured as the gas to dust ratio. Recently, the Planck survey has identified systematic variations in the submillimeter characteristics of interstellar dust, pointing towards the importance of dust evolution in the ISM (Planck Collaboration et al. 2014a). Changes in the size distribution of dust grains are expected from a theoretical perspective. Studies by Ormel et al. (2009, 2011) show that dust grains can grow in cold and dense molecular cloud, thus changing the size distribution. On the other hand, small dust grains and PAHs can be destroyed by the radiation field of stars. Furthermore, not only the size distribution but also the chemical composition of dust grains is expected to change with the environment (e.g. Henning 2010).

The extreme environments of regions of massive star formation may, in particular, be very conducive to dust processing. The study of dust around massive star forming regions has been of major interest since the beginning of modern infrared and radio studies. In recent years, the study of dust in HII regions has increased; mostly driven by the increase of observations at infrared wavelengths due to *Spitzer*, *Herschel* and most recently the Space Observatory for Infrared Astronomy (SOFIA). Deharveng et al. (2010), Anderson et al. (2012), Paladini et al. (2012) and Pellegrini et al. (2009) have studied the properties of dust in large evolved HII regions using combined datasets from *Spitzer*, *Herschel* and APEX. Arab et al. (2012) and Salgado et al. (2012) have revealed the importance of dust variations in individual HII regions. However, most of the data covered the far-infrared/sub-mm, as – until recently – *Spitzer*/MIPS was the only instrument but with limited angular resolution (6″) at mid-infrared wavelengths. Now that SOFIA is in operations, moderate spatial resolution (3–3.6″, depending on the wavelength) studies of dust emission from HII regions have come in reach. First results have been published on dusty HII regions Salgado et al. (2012); Hirsch et al. (2012); Lau et al. (2013, 2014). Here, we analyze data from the Faint Object infraRed Camera for the SOFIA Telescope (FORCAST) of the prototypical HII region, Orion (M42), from Shuping et al. (2012) more deeply to determine the characteristics of dust associated with the ionized gas and in the neutral photo dissociation region (PDR).

The Orion Nebula (M42, NGC1976) is the closest (414 pc, Menten et al. 2007) and the most extensively studied region of massive star formation in the Galaxy. The overall structure of the Orion nebula has been well studied at visible wavelengths (O'Dell 2001; O'Dell et al. 2008; O'Dell & Harris 2010). In the center of the nebula, four young massive stars known as the Trapezium have photoionized the surrounding gas creating an HII region; the global ionization structure is dominated by the most massive of the four stars, $\theta^1$ Ori C (spectral type $\sim$O7-O5V, Stahl et al. 2008; Kraus et al. 2007). The HII region is partially confined in a concave structure open towards the line of sight of the observer. A thin (0.05 pc) and dense ($\sim 10^5$ cm$^{-3}$) PDR is located between the background molecular cloud and the ionized gas. Due to the difference in pressure between the PDR and the ionized gas, material flows towards $\theta^1$ Ori C (O'dell & Bally 1999; O'Dell 2001). At approximately $2'$ south-east of the Trapezium the ionization front changes into an almost edge-on geometry producing the Orion Bar feature, the prototypical PDR (Tielens et al. 1993; Hollenbach & Tielens 1997). While several HII/PDR/molecular clouds have been studied, Orion is still a cornerstone in understanding the interaction of newly formed massive stars with the surrounding gas and dust.

Previous ground based studies of Orion in the mid-infrared regime have focused on the active star forming core, including the BN/KL object, and the region close to the Trapezium (Shuping et al. 2004; Robberto et al. 2005; Smith et al. 2005). These studies show a rich and complex structure, with diffuse extended emission, arcs, streaks and emission associated with proplyds. In addition, ground-based and space-based studies on the spatial variations of the unidentified infrared (UIR) bands – generally attributed to Polycyclic Aromatic Hydrocarbon molecules – have been reported by Cesarsky et al. (2000); Kassis et al. (2006); Galliano et al. (2008); Haraguchi et al. (2012). Recently, Arab et al. (2012) studied the far-infrared emission of Orion using *Herschel*/PACS and SPIRE photometry. They compared *Spitzer* and *Herschel* data and found a shift in distance between the Orion Bar as seen by *Herschel* and the IRAC 8 $\mu$m emission. Here, we will combine this data with the mid-infrared observations of the Orion HII region obtained by the FORCAST instrument (Herter et al. 2012) on board of Stratospheric Observatory for Infrared Astronomy (SOFIA).

In Section 2, we provide a description of the new FORCAST observations, as well as a summary for optical, infrared and submillimeter data from the literature used here. In Section 3, we give a description of the FORCAST images by comparing them with gas tracers. Together with *Herschel*/PACS far-infrared photometry, we construct spectral energy distributions (SED). The SEDs are further analyzed by using a two component modified blackbody model. In Section 4, the parameters derived from the SED fit are used to study the geometry of the Orion Bar and the properties of the dust grains in the ionized gas and the photodissociation region (PDR). In Section 5, we provide an explanation for the observed temperature of the dust in the ionized gas, as well as for the dust properties observed in Orion. We compare our results with previous works from the literature putting our results in the broader context of HII region evolution. Finally, in Section 6, we summarize our work and provide the conclusions.

## 2. Observations and Data

Two sets of observations were obtained by FORCAST on board of SOFIA during the Short Science flight series: the first one is centered in the Trapezium stars and has been presented in Shuping et al. (2012), with a field of view (FOV) of roughly $3' \times 3'$ the images cover most of the HII region in the center of Huygens nebula (Figure 1). The second set of images are centered in the Orion Bar with a similar FOV (Figure 1). The observations were made using FORCAST (Herter et al. 2012) on the 2.5 m telescope aboard SOFIA. FORCAST is a $256 \times 256$ pixel dual-channel, wide-field mid-infrared camera sensitive from 5 to 40 $\mu$m with a plate scale of $0.768''$/pixel and field of view of $3.4 \times 3.2'$. The two channels consist of a short wavelength camera (SWC) operating from 5 to 25 $\mu$m and a long wavelength camera (LWC) operating from 28 to 40 $\mu$m. An internal dichroic beam-splitter enables simultaneous observation from both long and short wavelength cameras. A series of band pass filters are used to image at selected wavelengths.

The Orion Bar data was taken over three flights: 01-Dec-2010, 04-Dec-2010 and 08-Dec-2010 with FORCAST. Dichroic mode was used to simultaneously observe in a number of filters: 19.7/31.4, 19.7/37.1, 7.7/37.1, 6.6/37.1, and 11.3/37.1. Direct (single channel) mode was also employed for filters: 6.4, 6.6, 7.7, 11.3 and 37.1 microns. A southern region of the Orion Bar was also observed in the dichroic mode with the 19.7/31.4 and 19.7/37.1 pairs. Data were reduced and calibrated as discussed in Herter et al. (2013).

A small rotation is seen in our Orion Bar centered data for the 7.7 and the 31.7 $\mu$m images, when comparing to the Trapezium centered images. Since no obvious point sources are detected in most of our images, another procedure must be performed to correct for this effect. At each wavelength the images are rotated in discrete steps of $0.1°$ from $-4°$ to $4°$ and, by using a normalized cross-correlation estimator, a best expected rotation angle is obtained. The process was performed for all of our images, but only for the 7.7 and 31.7 $\mu$m images is the rotation significant ($2°$). We checked the astrometry of our data after the correction is applied by comparing the 7.7 $\mu$m to the 8 $\mu$m *Spitzer*/IRAC image and the agreement is good to within about $1''$. Finally, the images were combined by averaging the flux in the region of overlap and scaling the full images accordingly.

We checked the relative flux calibration of the FORCAST images by comparing them with ISOCAM-LWS spectra (Cesarsky et al. 2000). The agreement is good in all bands. Using the ISOCAM-LWS spectra we estimate the contribution from the [S III] 18.7 $\mu$m line to the 19.7 $\mu$m flux to be less than 10 % at the ISOCAM-LWS positions.

Images from the PACS instrument (Poglitsch et al. 2010) on board of the *Herschel Space Observatory* (Pilbratt et al. 2010) at 70 $\mu$m and 160 $\mu$m were obtained by Abergel et al. (2010) as a part of the "Evolution of interstellar dust" key program and have been presented in Arab et al. (2012). Centered at the Orion Bar and with a FOV of about $14 \times 15'$ the images overlap with our FORCAST images. Including the *Herschel*/PACS images allows us to analyze the full dust spectral energy distribution (SED) at infrared wavelengths. The spatial resolution of the images is $5.6''$ and $11.3''$ at 70 $\mu$m and 160 $\mu$m, respectively.

Spectral cubes of infrared cooling lines [O I] at 63 $\mu$m 145 $\mu$m and [C II] at 157 $\mu$m were observed by PACS and were published by Bernard-Salas et al. (2012). The spatial resolution of this data changes with wavelength being 4.5″ at 63 $\mu$m 10″ at 145 $\mu$m and 11″ at 157 $\mu$m. The FOV is smaller than our FORCAST images overlapping only in the Orion Bar.

Since the mid-infrared and far-infrared emission probes the dust emission, we complement our dataset with low-resolution (10 Å FWHM) optical spectroscopy from the Postdam MultiAperture Spectrograph (PMAS) in the PPAK mode at the 3.5m telescope in Calar Alto Observatory (Sanchez et al. 2007). The data set consists of ∼8000 spectra in the 3700-7000 Å range, centered in the Trapezium stars and covering 5′×5′. Although the data was taken under non-photometric conditions, the spatial resolution is good enough for a spatial comparison with our infrared data. The astrometry was cross calibrated by comparing the data with Hubble Space Telescope images observed with WFPC.

We extracted the integrated flux and peak emission for some spectral lines at each pixel position, thus we created 2-D maps for the most important lines that trace gas physical properties such as H$\alpha$, H$\beta$, [S II]$\lambda$6716,6731, [O II]$\lambda$3727,3729, [O III]$\lambda$4363, [O III]$\lambda$4959,5007, and the [O I] at 6300Å. In our analysis, we studied the spatial distribution of [O I] as a tracer of the ionization front, we neglect possible contamination due to sky emission [1].

Rotational transition lines of $^{13}$CO and C$^{18}$O in the $J = 3 \rightarrow 2$ transition were observed by the James Clerck Maxwell Telescope (JCMT) using HAARP/ACSIS, published by Buckle et al. (2012). The resolution is close to 15″ and the field of view (FOV) overlaps with our dataset.

Simón-Díaz et al. (2006) studied the spectra of eight stars including $\theta^1$ Ori C. They determined a $T_{eff}$ = 39000 ± 1000 K and a log $g$ = 4.1 corresponding to an O5.5V-O6V star in the scale of Martins et al. (2005). By interpolating the values of Table 1 in Martins et al. (2005), we derived a stellar luminosity $L = 2.23 \times 10^5$ L$_\odot$ and number of ionizing photons $N_{ion} = 1.06 \times 10^{49}$ s$^{-1}$. The second most massive star is $\theta^1$ Ori D a B0.5V star Simón-Díaz et al. (2006), but with an effective temperature more similar to an O9 star. The star $\theta^1$ Ori A is a B0.5 star. Using the values for the stars we integrated the spectra of the stellar models (Martins et al. 2005 and Kurucz 1993 for B0.5V stars) from 912 Å to 1000 Å to obtain a total FUV luminosity in the range $1.4-2.7 \times 10^5$ L$_\odot$. The properties of the Trapezium stars are summarized in Table 2.

---

[1] The [O I]$\lambda$6300Åflux is expected to be only a 20% larger than that of the sky and, due to the low spectral resolution of the PMAS instrument, both lines are not easily separable. However, the contribution from the SKY emission, for the purpose of the analysis presented here, can be treated as random noise (or rather a baseline contribution to the emission). The emission at the positions of interest for our analysis is well above the $3\sigma$ confidence level.

## 3. Analysis

### 3.1. An Overall View of the Orion Nebula

The Orion Nebula is one of the most studied massive star forming regions. At optical wavelengths, the gas in the Orion Nebula shows a complex structure (e.g. Doi et al. 2004) In addition to stars, diffuse emission, the ionization front, as well as proplyds and jets have been thoroughly studied. As can be seen in Figures 1 and 2, the FORCAST mid-infrared images show an intricate morphology that changes with wavelength. In particular, the 6.4, 6.6 and 7.7 $\mu$m images (PAH images) show the Orion Bar, the Trapezium stars and the Ney-Allen nebula. In the 19 $\mu$m band, in addition, the Herbig-Haro object HH203, is clearly visible and there is also emission associated with the stars, $\theta^2$ Ori A and $\theta^2$ Ori B. The BN/KL object (north of the Trapezium), the brightest region in the FORCAST images, is saturated in the *Herschel*/PACS bands.

Close to the Trapezium stars and towards the south-west there is bright extended emission in all PAH and mid-infrared images, this emission seems co-spatial with the peak in emission in radio and H$\alpha$ images (Dicker et al. 2009). south-east of the Trapezium, extended emission and substructure is seen as "filaments", arcs, and "ripples" towards the Orion Bar in all the FORCAST images. We note that this spatial structure is not correlated with the optical extinction maps derived from H$\alpha$/H$\beta$ maps (O'Dell & Yusef-Zadeh 2000; Sanchez et al. 2007), which is consistent with the idea that most of the extinction towards the region is produced by a foreground cloud. Instead, these structures are thought to be perturbations in the surface of the background PDR (Shuping 2012).

The high resolution and large FOV of the FORCAST images together with the large amount of observations from the literature allow us to directly compare the spatial distribution of dust emission with different gas tracers. However, a close comparison between the IR and optical maps is hampered by the presence of emission associated with the PDR and HH203/204 jets penetrating into the ionized volume. Comparison of the 19.7 and 7.7 $\mu$m maps reveals similarity for some spatial structures which we interpret as being associated with the PDR. Nevertheless, several regions bright in 19.7 $\mu$m emission do not show up in the 7.7 $\mu$m image. Hence, we consider that this part of the 19.7 $\mu$m emission is associated with dust in the ionized gas rather than in the background PDR. To bring this out, we have subtracted the 7.7 $\mu$m map from the 19.7 $\mu$m map scaled to the intensity ratio in the Orion Bar PDR, the result is shown in Figure 3. Perusing the velocity dependent H$\alpha$ maps (Doi et al. 2004), we recognize a resemblance to several structural features in the subtracted 19.7 $\mu$m emission with the H$\alpha$ emission in the range of 0-8 km s$^{-1}$ (with respect to LSR, Doi et al. 2004) particularly directly north of the Bar. However, not surprisingly, there is no one-to-one correspondence as gas and dust emission processes differ in their density and temperature dependence. Moreover, we also recognize that decoupling of gas and dust seems to be a common phenomena in blister HII regions (Ochsendorf et al. 2014b) and a one-to-one correspondence is not expected. We also notice that some of the dust emission can be related to the "Big Arc", in the terms of Doi et al. (2004). Furthermore, the 19.7 $\mu$m emission is not related to the molecular cloud (cf. the 70 and 160 $\mu m$, Figure 4). While some of the emission can be related to the PDR (as traced by PAH emission, Figure 4), the rest of the emission seem to be associated with ionized gas (cf. Figure 2 and, for instance, O'Dell & Doi (2003)).

At longer wavelengths, the PACS images show the Orion Bar as a bright feature, but the brightest emission is associated with the Orion molecular cloud (OMC-1). Note that, unlike the PAH and mid-infrared emission, the region close to the Trapezium stars does not show strong, associated, far-infrared emission in the PACS bands. By comparing the far-infrared with the optical images, the only similarities are found towards the ionization fronts in the direction of the Orion Bar and Orion S.

To further analyze the spatial distribution of the emission, we extracted cross-cuts starting from $\theta^1$ Ori C to deep behind the Orion Bar at the position shown in Figure 1. The extracted profiles are shown in Figure 4. A general trend is that the dust emission is bright close to the Trapezium and drops with distance up to 0.19 pc where the Orion Bar is prominent (see Section 3.2). A closer look reveals the presence of small structures in all the FORCAST bands at $10''$, $30''$ and $80''$, these structures are more prominent at 19.7 $\mu m$.

### 3.2. The Orion Bar

All images show a prominent structure to the south-west associated with the well known ionization front, the Orion Bar (Figures 1 and 2). Yet, there are subtle variations in the maps. Specifically, the position of the peak emission changes with wavelength (Table 3). Indeed, comparison of the H$\alpha$, [O I] 6300Å, PAHs, H$_2$ and $^{13}$CO emission reveal the layered structure expected for an edge-on PDR (Figure 4; Tielens et al. 1993). The dust emission show a similar stratified structure, with shorter wavelengths peaking closer to the Trapezium stars as compared with longer wavelength images, consistent with heating by the Trapezium stars of an edge-on slab (Werner et al. 1976; Tielens et al. 1993; Arab et al. 2012). The ionization front, as traced by the [O I] 6300 Å profile, show a sharp rise and a "wing-like" decay, but the peak is located closer to $\theta^1$ Ori C at $113''$ (0.23 pc). The H$\alpha$ peak emission is located even closer to $\theta^1$ Ori C at $109''$ (0.22 pc). The PAH, 31 $\mu m$ and 37 $\mu m$ images show profiles with a sharp rise and a peak at $118''$ (0.24 pc). The 19.7 $\mu m$ image shows, in addition, a separate peak associated with the HII region but there is no clear counterpart in any other ionized optical line. At longer wavelengths, the 70 $\mu m$ also show a sharp rise at the bar position but the peak is broader than at mid-infrared wavelengths. This broadening trend seems to continue in the 160 $\mu m$ data, but this trend is clearly an effect of the lower resolution of PACS at 160 $\mu m$. The $^{13}$CO show a broader "gaussian-like" profile with peaks at $138''$ (0.28 pc), respectively. At 450 $\mu m$, the peak of the emission has shifted inwards and occurs at the $^{13}$CO peak[2]. The ground-based 850 $\mu m$ data, on the other hand, shows an emission peak that encompasses both the mid-infrared and the 450 $\mu$m peak (Figure 4). The "shoulder" seen in the 850 $\mu$m images at distances less than 0.2 pc is, most likely, free-free emission produced in the ionized gas.

---

[2] Some negative values were found in the 450 $\mu m$ image we added a constant to the data.

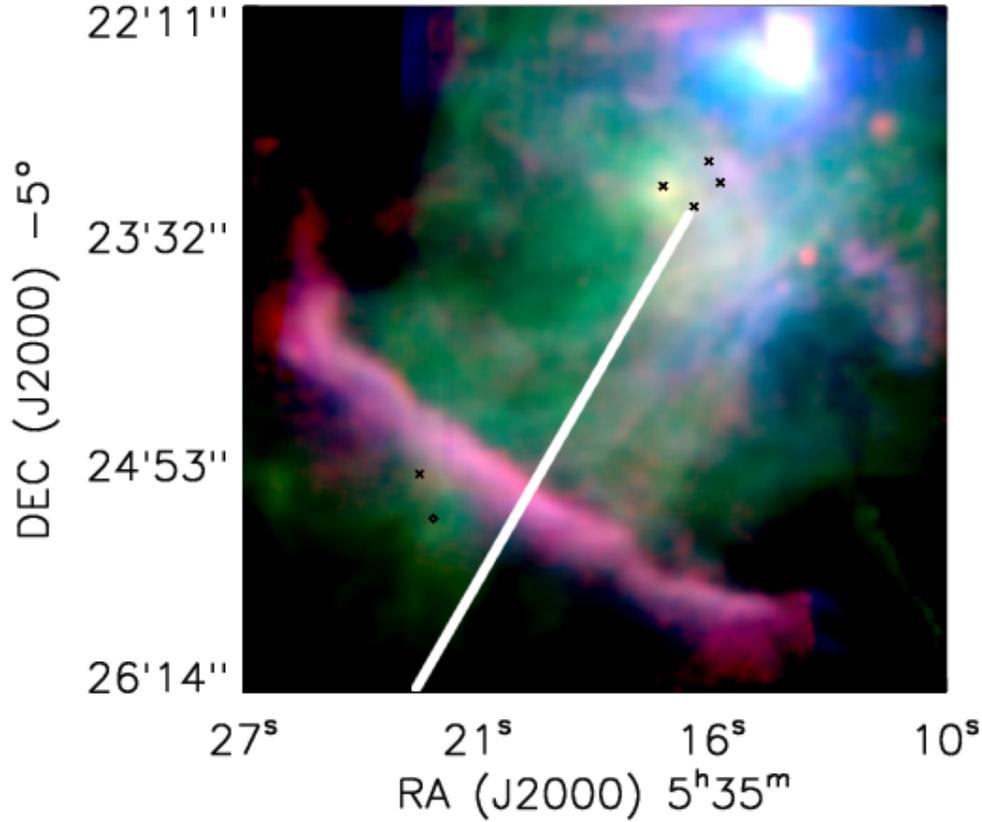

Fig. 1.— False color image for the FORCAST central region (blue 7.7, green 19.7, red 37.1). The Trapezium stars are marked with crosses. The Orion Bar is prominent in the south-east while the BN/KL region dominates the north-west (see text for details). The white line show the cross-cut shown in Figure 4.

Assuming a constant PDR density of $10^5$ cm$^{-3}$ (Simon et al. 1997), we can quantify the location of the different emission peaks in terms of the total column density. We locate the surface of the PDR with the emission peak of the PAH features, at $117.5''$ from $\theta^1$ Ori C  (see Table 3). The mid-infrared dust continuum peaks slightly deeper in to the PDR. The 160 $\mu$m emission shows a rather broad peak, running from $120''$ to $140''$, while the sub-millimeter emission peaks at about $140''$. For comparison, the $^{13}$CO peak is located at a column density of $N_{\rm H} = 12.7 \times 10^{21}$ cm$^{-2}$, while the peak in H$_2$ emission is located at $20''$ from the ionization front, roughly 0.04 pc or at a depth $N_H = 7.74 \times 10^{21}$ cm$^{-2}$. At a distance of 200$''$ (0.4 pc), the emission in all the bands has dropped to less than 20% of the peak emission at the Bar. The sub-millimeter data conclusively show that the Orion Bar is limited in extent in to the molecular cloud to $\simeq 140''$ or a total column of $N_H = 13.9 \times 10^{22}$ cm$^{-2}$. The Orion Bar is a shell bounded on one side by the ionization front and on the other side (likely) by the shock front running into the molecular cloud.

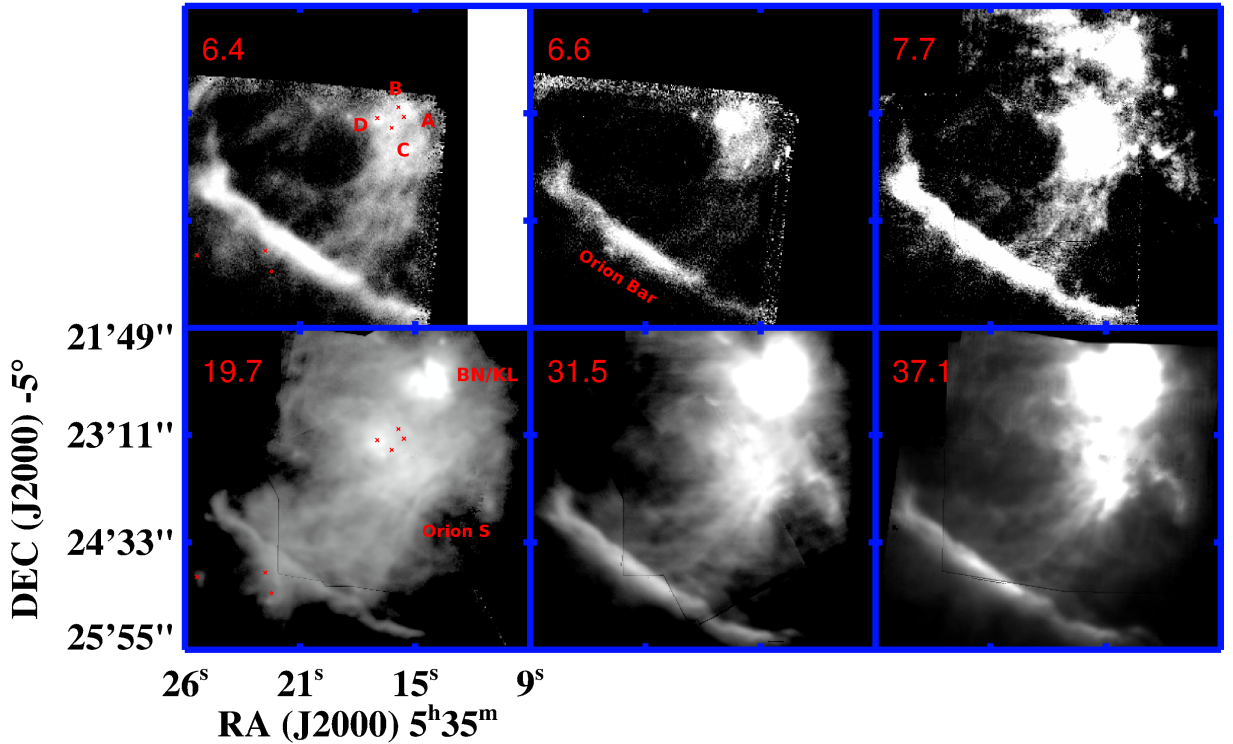

Fig. 2.— The complete data set of FORCAST images at 6.4, 6.6, 7.7, 19.7, 31.5 and 37.1 $\mu$m. Red crosses correspond to the Trapezium stars and $\theta^2$ Ori A and $\theta^2$ Ori B. The diamond marks the location of HH 203/204.

In neutral regions the cooling is dominated by three forbidden transitions lines: [O I] at 63 $\mu$m [C II] at 157 $\mu$m and [O I] at 145 $\mu$m. The [O I] 63 $\mu$m and the [C II] 157 $\mu$m line peaks are located at a distance of 125$''$ (0.25 pc) corresponding to a column density of $N_H = 4.65 \times 10^{21}$ cm$^{-2}$. This is slightly deeper in the Orion Bar than the PAH and mid-infrared emission peaks (Figure 4). We also note that clumps seen in the [O I] and [C II] IR lines by Bernard-Salas et al. (2012) are also evident in mid-infrared continuum emission. The [O I] and [C II] structures are, however, located somewhat deeper than clumps in the dust and PAH emission maps (Figure 5).

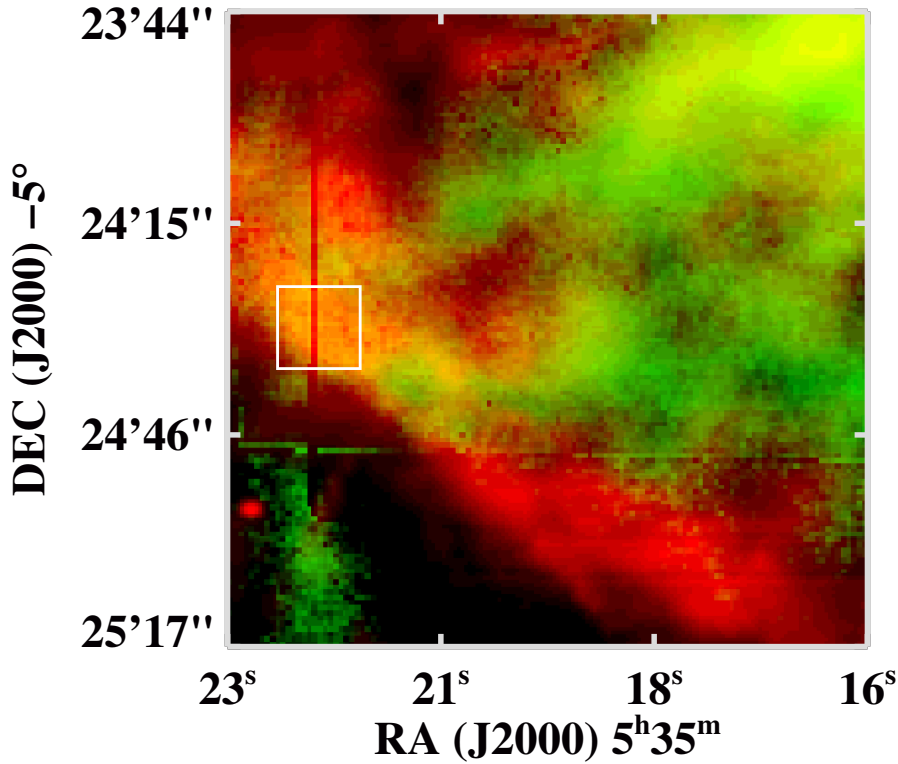

Fig. 3.— A comparison between the 19 $\mu$m PDR-subtracted emission (green) and H$\alpha$ at 8 km/s from (Doi et al. 2004) (red). While there is no one-to-one correspondence between the two images, some features are co-spacial. The white box marks one of the regions where there is correspondence between dust and gas emission.

### 3.3. Spectral Energy Distribution Fitting

The SOFIA/FORCAST data provide, for the first time, full spectral coverage of the SED of the dust emission in the Orion Bar with high angular resolution. In fact, the FORCAST filters are well matched to the peak of the SED, which shifts from about 20 $\mu$m to 70 $\mu$m over this region. In order to analyze the SED of our images, the FORCAST and the PACS 70 $\mu$m images were convolved to the PACS 160 $\mu$m resolution assuming a Gaussian PSF and using the values in Table 1 for the FWHM. The images were regridded to a common pixel scale of 6.4″, the native resolution of PACS at 160 μm. Figure 6 shows SEDs at a few selected positions. The SED peak changes with position from around 30 μm close to the Trapezium and moving to 80 μm behind the Orion Bar.

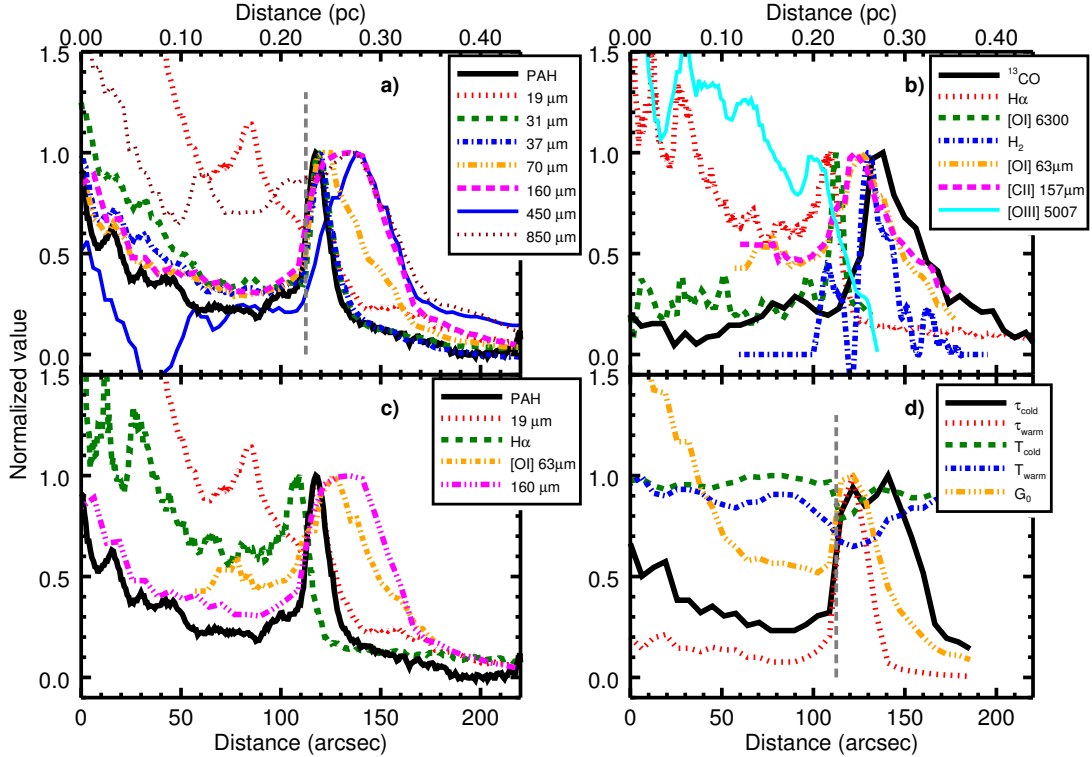

Fig. 4.— Cross-cuts of the Orion Bar in different tracers (the cross-cuts were extracted at the native resolution of the different instruments involved). This diverse data set unveils the onion-skin nature of the PDR. The SOFIA/FORCAST data traces warm dust and PAHs. Optical Hα traces ionized gas while the [O I] at 6300 Å locate the ionization front (shown as a gray-dashed line in the other panels). Molecular gas in the PDR is traced by the ¹³CO J=3-2 line (Buckle et al. 2012) and the H₂ 1-0 S(1) (van der Werf et al. 1996). The extent of the cold dust in the Orion Bar is revealed by the 450 μm and 850 μm JCMT SCUBA data (Johnstone & Bally 1999).

The emission seen at wavelengths shorter than ∼ 10 μm is expected to be produced by PAHs and very small grains in the background PDR and in the molecular cloud. Furthermore, The ISOCAM-CVF spectra (Cesarsky et al. 2000) clearly shows the PAH bands at 6.4 and 7.7 μm. Since PAHs and very small grains are stochastically heated, therefore we decided to exclude the emission at these wavelengths from the SED analysis. At wavelengths larger than ∼ 10 μm, the SEDs are very broad and show evidence for two separate components; particularly beyond the Orion Bar. Indeed, modified blackbody fits to the SEDs at each pixel position with a spectral index β = 1.8 are not satisfactory. Modified blackbody fits with variable β (between 0 and 3) yield better results in terms of the reduced $\chi^2$. However, the $\beta$ values are too low ($\beta < 0.5$), far from standard values.

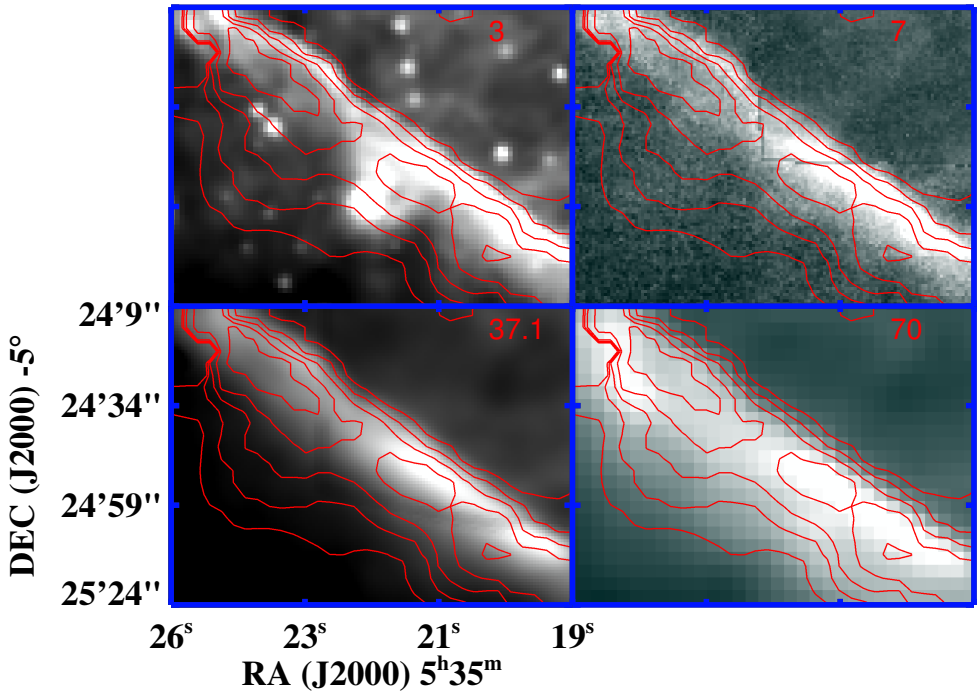

Fig. 5.— *Upper Left Spitzer*/IRAC 3 $\mu$m image at the Orion Bar. *Upper Right* FORCAST 7.7 $\mu$m image at the Orion Bar. *Lower Left* FORCAST 37.1 $\mu$m image at the Orion Bar. *Lower Right* PACS 70 $\mu$m image at the Orion Bar. The red contours are the [O I] 63 $\mu$m integrated emission, the peak of the line emission is located deeper into the PDR as compared to the dust tracers.

The IR emission observed in the maps results from contribution of grains located within the HII region, in the PDR and in the molecular cloud. When comparing the PACS 160 $\mu$m emission with the $^{13}$CO data, it is clear that there is a component associated with the Orion molecular cloud (see Figure 7). The mid-infrared emission observed by SOFIA/FORCAST and far-infrared emission from *Herschel*/PACS 70 $\mu$m are tracing different components of the dust. This can clearly be seen at positions close to the Orion Bar where the combination of two modified-blackbody function are needed to obtain a good description of the data (see Figure 6 for examples of the SEDs and fits). Therefore, we fitted all the data with two modified blackbody components assuming optically thin emission using: $I_\lambda = \tau_{warm} \times B_\lambda(T_{warm}) \times (\lambda/\lambda_0)^{-\beta} + \tau_{cold} \times B_\lambda(T_{cold})(\lambda/\lambda_0)^{-\beta}$ , where warm corresponds to $T_{warm} > 70\ K$ and cold to where $T_{cold} < 70\ K$. In order to limit the number of free parameters, we have fixed the exponent of the modified blackbody for both components ($\beta = 1.8$, Planck Collaboration et al. 2011). The constants, $\tau_{warm}$ and $\tau_{cold}$, are the optical depth at a given reference wavelength (we adopt $\lambda_0 = 100\ \mu$m).

Table 1: FORCAST and PACS angular resolution

| Wavelength ($\mu$m) | FWHM ($''$) |
|---|---|
| 6.4 | 2.9 |
| 6.6 | 3.0 |
| 7.7 | 2.7 |
| 19.7 | 2.9 |
| 31.5 | 3.4 |
| 37.1 | 3.6 |
| 70 | 5.6 |
| 160 | 11.3 |

Table 2: Properties of the central stars (Martins et al. 2005)

| Star | spectral type | Ionizing photon flux | Luminosity log(L/L⊙) | FUV Luminosity log(L/L⊙) |
|---|---|---|---|---|
| $\theta^1$ Ori A | B0.5V | 47.36 | 4.52 | 4.61 |
| $\theta^1$ Ori B | B1V | 47.06 | 4.42 | — |
| $\theta^1$ Ori C | O4V-O6V | 48.96 | 5.68-5.31 | 5.26 (O5.5) |
| $\theta^1$ Ori D | B0.5V | 47.36 | 4.52 | 4.61 |

Dust grains can be either in equilibrium with the radiation field or stochastically heated by photons depending on the size of the grain and the strength of the radiation field. In Orion, due to the strength of the radiation field ($\sim 1 \times 10^5\ G_0$, Section 4.1), the emission of dust particles is produced by dust particles in radiative equilibrium with the radiation field, irrespective of their size (except for PAHs). Indeed, small (carbonaceous) dust particles of $\gtrsim 10$ Å in size ($\gtrsim 120$ C atoms) are expected to be in equilibrium with the radiation field (Draine & Li 2001, 2007). The steep increase in density (and extinction) at the Orion Bar produces a clear broadening of the SEDs. While large grains remain in equilibrium with the (lower) radiation field, small grains are stochastically heated. Therefore, deeper in the Orion Bar PDR, the warm component is produced by small grains stochastically heated.

Figure 8 shows the temperature and optical depth maps for both components at the positions where our fit is reasonably good (reduced $\chi^2 < 3$). Clearly, PACS 160 $\mu$m trace a cold dust component, while the FORCAST mid-infrared images trace a warm dust component. The contribution to the PACS 70 $\mu$m emission is a combination of both the warm and the cold component (see also Figure 9). The cold dust component reveals a temperature in the 38-44 K range to the east and south-east of the Trapezium. At the Orion Bar the temperature decreases to values of about 33 K.

Table 3: Spatial emission peaks

| Wavelength ($\mu$m)/Line | Distance ($''$)[a] | $N_{\mathrm{H}}$[e] ($\times 10^{21}$cm$^{-2}$) | comment |
|---|---|---|---|
| 6.4 | $117.5 \pm 0.8$[b] | 0 | PDR surface |
| 6.6 | $117.5 \pm 0.8$ | 0 | |
| 7.7 | $117.5 \pm 0.8$ | 0 | |
| 19.7 | $118 \pm 0.8$ | 0.31 | |
| 31.5 | $119.5 \pm 0.8$ | 1.24 | |
| 37.1 | $120 \pm 0.8$ | 1.55 | |
| 70 | $125 \pm 3$ | 4.65 | |
| 160 | $135 \pm 6$ | 10.8 | |
| 450 | $138 \pm 6$ | 12.7 | |
| 850 | $138 \pm 6$ | 12.7 | |
| [O I]$_{63}$ | $125 \pm 10$ | 4.65 | |
| [C II]$_{157}$ | $125 \pm 10$ | 4.65 | |
| [O I]$_{6300}$ | $112.5 \pm 3$ | ... | Ionization front |
| H$_\alpha$[c] | $109 \pm 0.2$ | ... | Ionization Bar |
| H$_2$[d] | 130 | 7.74 | Peak of the 1-0 S(1) |
| $^{13}$CO(3-2)[f] | $138 \pm 6$ | 12.7 | Peak of low-$J$ $^{13}$CO |

[a] Measured from $\theta^1$ Ori C.

[b] Errors are the pixel scale of the images.

[c] Taken from HST images O'Dell & Yusef-Zadeh (2000).

[d] Maps from van der Werf et al. (1996).

[e] $N_{\mathrm{H}}$ measured from the PDR surface, using the spatial extent and adopting a density of $1 \times 10^5$ cm$^{-3}$ (Simon et al. 1997).

[f] Taken from Buckle et al. (2012).

Towards Orion S and Orion BN/KL the temperature is also low. Unfortunately, towards Orion BN/KL the PACS images are saturated and the fit results are not reliable. In terms of the optical depth, the warm component shows a spatial correlation with the ionized gas emission and is fairly uniform (Figure 8c). An increase in $\tau_{warm}$ is seen towards Orion S, the Orion Bar and the BN/KL object. Variations in temperature for the warm component are large, ranging from 70 K at the Orion Bar to values larger than 100 K in the ionized gas. Close to $\theta^1$ Ori D, the temperature of the Ney-Allen nebula reaches values of 120 K. A high temperature (110 K), as compared to its surrounding area, is seen towards the Herbig-Haro object HH 203. The peak in $\tau_{warm}$ is located at the same distance as the PAH/mid-infrared peak ($120''$), while the cold dust component show two peaks at $120''$ and $140''$.

Perusal of the optical depths maps reveals that the cold component is strongly concentrated towards the Orion Molecular Cloud Core OMC-1, and besides the BN/KL region we also recognize

Orion S, but the Orion Bar does not figure prominently. In contrast, the Orion Bar is very evident in the optical depth map of the warm component. The temperature map provides a complementary view: The cold component reaches a particularly low temperature in the cold molecular cloud core – as evident in the temperature map of the cold component – as expected from the high column density associated with this feature. We also note that the highest temperatures for the cold component are present in the ionized gas towards the west of the Trapezium and also within a small pocket near $\theta^2$ Ori A. This same area around $\theta^2$ Ori A and HH 203 sticks out as particularly warm in the warm component map. At first sight surprisingly, the Orion Bar shows up in the temperature map of the warm component as a relatively cold feature compared to the much warm material associated with the ionized gas. This reinforces the discussion in Section 3.1 that much of the mid-IR emission towards the north of the Orion Bar is associated with dust in the ionized gas. We also note the increased presence of the warm dust component beyond the extent of the Orion Bar traced in the submillimeter, which we attribute to dust associated with a thin layer of ionized gas that is quite evident in the optical and IR spectroscopic data (Rubin et al. 2011; Boersma et al. 2012).

The total dust emission was estimated by integrating the flux from the modified-blackbody fits. We can also assess the contribution from the cold and warm component to the overall IR emission. In the ionized zone, towards the Trapezium stars, the total luminosity is dominated by the warm component, being a factor of seven larger than the cold component on average. In the HII region, the ratio decreases with distance up to $115''$ – the Orion Bar – where it suddenly increases. Beyond the surface of the Orion Bar, the ratio decreases again, reaching a minimum of 0.6 at $\sim 150''$. The further increase beyond this minimum (towards the south-east) is not associated with the bar as outlined by the sub-millimeter emission.

As mentioned in Section 3.1, the infrared emission is dominated by the BN/KL object, a region around the Trapezium and the Orion Bar PDR. We focus on the Trapezium region and the Bar, as the BN/KL object is not covered in all of our images. We extracted the total luminosity in the region shown in Figure 9: one in a rectangular aperture of $158 \times 20''$ ($0.32 \times 0.04$ pc$^2$) at the Orion Bar. The total luminosity at the surface of the Orion Bar is $9.5 \times 10^3$ L$_\odot$.

## 4. Results

### 4.1. Geometry of the Orion Bar

The observed total infrared emission is a measure for the amount of stellar energy intercepted by the Orion Bar PDR and we can use this value to determine the geometric dimensions of the Bar. We will use the emission traced by the warm dust component to estimate the energy absorbed by the dust in the Orion Bar region ($L_{IR} = 9.5 \times 10^3$ L$_\odot$). We can derive a value for the absorbing area at the surface of the PDR, i.e.: $A_{abs} = L_{IR}/\left(L_\star/4\pi\, d^2\right)$. The minimum stellar luminosity of $1.4 \times 10^5$ L$_\odot$ corresponds to the often adopted $G_0$ of $5 \times 10^4$ (Tielens & Hollenbach 1985), while the twice higher luminosity of the Trapezium cluster (Section 2) results in $G_0$ that is double this value. In contrast, Marconi et al. (1998) derived a $G_0$ of $2.6 \times 10^4$ from the fluorescent, near-IR OI lines, originating in the ionization front. Adopting this latter value, the absorbing area is $0.09 \pm 0.02$ pc$^2$. Adopting a rectangular geometry and using the measured size of the Orion Bar (0.32 pc) as the other side of the rectangle, we obtain a transverse size of $l_{LOS} = 0.28 \pm 0.06$ pc.

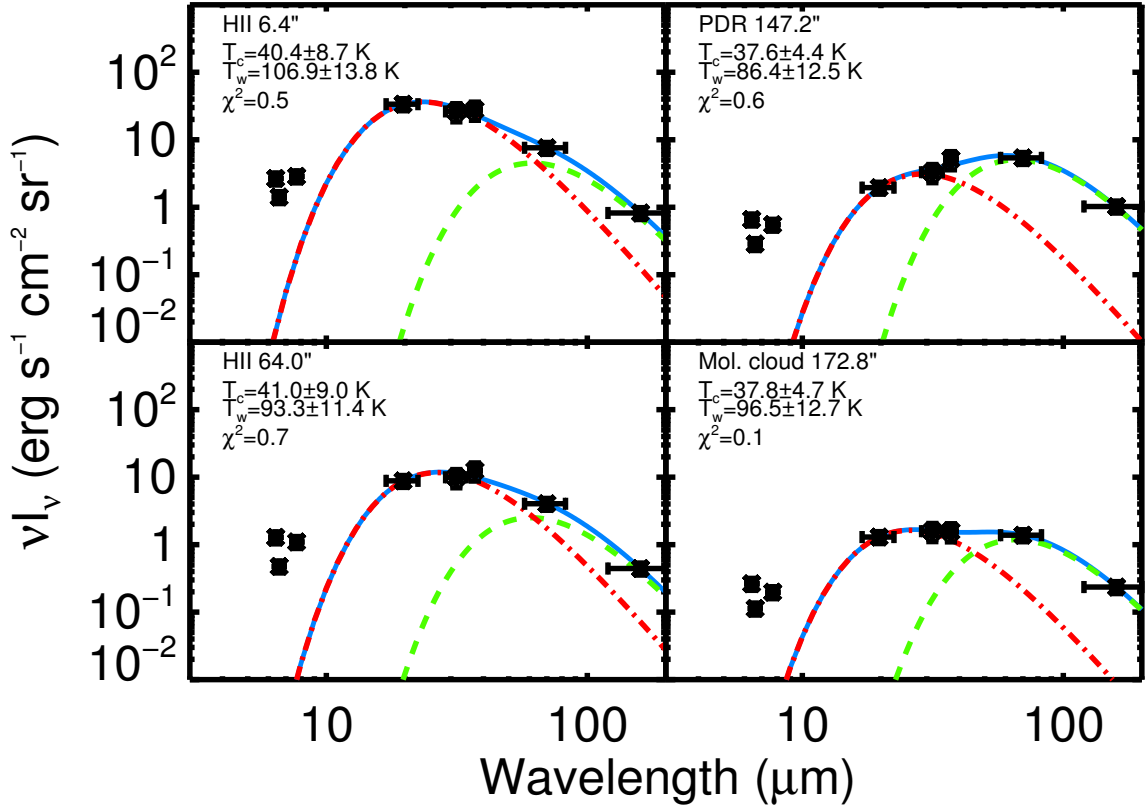

Fig. 6.— Examples of the SEDs at four positions. The results of the modified blackbody fits are also plotted.

The line of sight extent can also be derived from observed atomic or molecular column densities in the Orion Bar PDR, adopting a gas density. The observed C$^+$ and C$^{18}$O column densities transformed to total H column densities using standard gas phase abundance ratios ($X(\text{C}^+) = 1.5 \times 10^{-4}$ and $X(\text{C}^{18}\text{O}) = 1 \times 10^{-7}$; Cardelli et al. 1996; Tielens 2005) are $7 \times 10^{22}$ and $1.3 \times 10^{23}$ cm$^{-2}$, respectively (Hogerheijde et al. 1995; Ossenkopf et al. 2013). The density in the PDR has been obtained from modeling molecular emission lines of $10^5$ cm$^{-3}$ (Simon et al. 1997). Slightly lower densities have been obtained from an analysis of the PDR cooling lines and from the observed stratification of the Bar adopting standard dust properties (Tielens et al. 1993) but we deem those more uncertain. Adopting the $C^+$-derived H-column density as appropriate for the atomic zone in the PDR, and a density of $1 \times 10^5$ cm$^{-3}$, we derive a length-scale of 0.23 pc for the line of sight. For comparison, deriving the EM associated with the ionization bar from the observed H$\alpha$ intensity (Figure 4 in Wen & O'Dell 1995) and the density derived from the [S II] lines ($2.5 \times 10^3$ cm$^{-3}$, Sanchez et al. 2007), we also arrive to a line of sight extent of the (ionized) bar of 0.3 pc. Note, though, that there is a large column density of foreground ionized gas in the same direction.

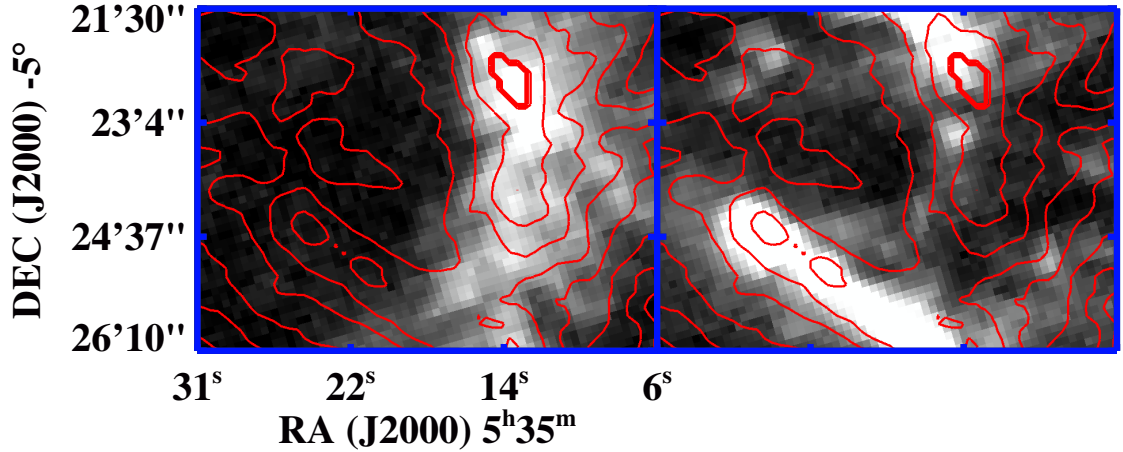

Fig. 7.— $^{13}$CO data at two velocity components corresponding to OMC (left) and Orion Bar (right), the red contours are the PACS 160 $\mu$m emission.

In summary, adopting the Marconi et al. (1998) value for $G_0$ measured in the ionization front, we arrive at $l_{LOS} = 0.28$ pc. This then implies that the actual distance from the Trapezium stars to the Orion Bar is $0.33 - 0.45$ pc – depending on the adopted luminosity – rather than the projected distance of 0.23 pc. Finally, from the observed width of the [O I] 6300 Å structure (0.02 pc) – much larger than the expected width of an ionization front ($10^{-3}$ pc Tielens 2005) –, and the derived length scale of the Orion Bar, we arrive at an inclination of the Orion Bar of 4° with respect to the line of sight.

### 4.2. The Swept-up Shell

In Section 3.2, we identified the extent of the Orion Bar on the sky as the location of the shock front, some 0.1 pc from the ionization front. Most of the gas in this swept-up shell will be molecular and in pressure equilibrium with the atomic gas ($n = 10^5$ cm$^{-1}$, $T = 500$ K; Simon et al. (1997); Allers et al. (2005)) in the PDR. Adopting a temperature of 50 K for the molecular gas, we then arrive at a density of $10^6$ cm$^{-3}$. The radial column density of the bar is then $N_H \simeq 3 \times 10^{23}$ cm$^{-2}$ and the total mass of swept-up material is $6 \times 10^3$ M$_\odot$. As the mass of ionized gas is very small ($\simeq 10$ M$_\odot$, we can infer an average density of $3 \times 10^4$ cm$^{-3}$ before the star turned on; a typical density for a massive molecular cloud core (e.g.: Rathborne et al. 2006; Liu et al. 2014). Comparing this density with that of the ionized gas, we arrive at a shock velocity of $v_s \simeq \sqrt{n_o/n_e}\, C_s \simeq 3$ km s$^{-1}$ with $C_s \approx 10$ km s$^{-1}$, the sound velocity in the ionized gas, and $n_o$ and $n_e$ the densities of the neutral and ionized material, respectively (Tielens 2005). Coupled with the location of the shell, we infer an age of $10^5$ yr for $\theta^1$ Ori C and the formation of the HII region. This should be considered as an upper limit as the shock may have decelerated with time. Using the size of a "measurable" HII region to infer the age of the ionizing star should be considered as an order of magnitude estimation, since this heavily depends on the structure and density distribution of the natal environment as well as the viewing angle of the observed system. Our value, however, is similar to that estimated by Hillenbrand (1997) for the age of low-mass stars in the region and much less than that of Simón-Díaz et al. (2006) (2.5±0.5 Myr).

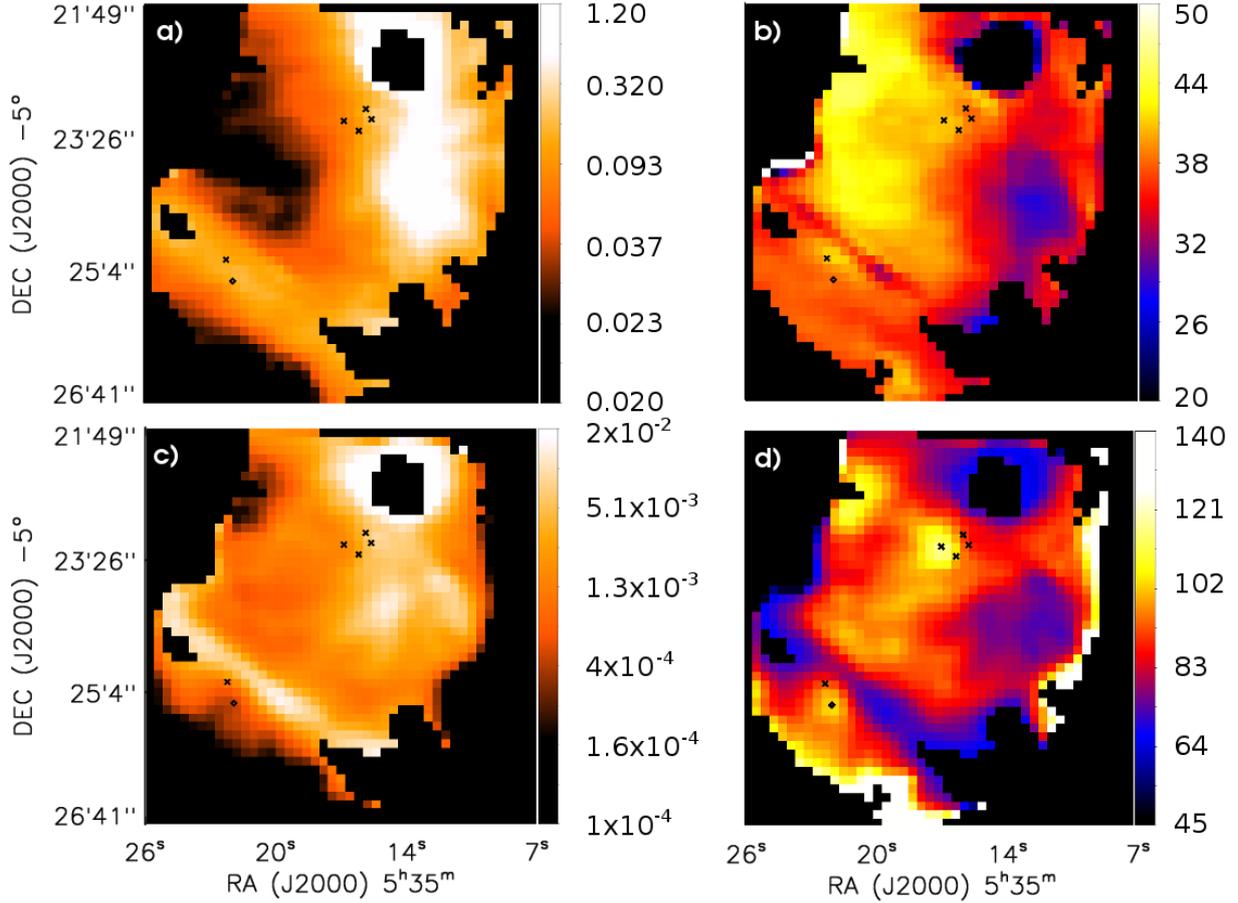

Fig. 8.— Temperature and optical depth maps of the two dust components revealed by the SED modeling. a) Optical depth,$\tau_{cold}$, for the cold component. b) Temperature map for the cold component. c) Optical depth, $\tau_{warm}$, for the warm component. d) Temperature map for the warm component.

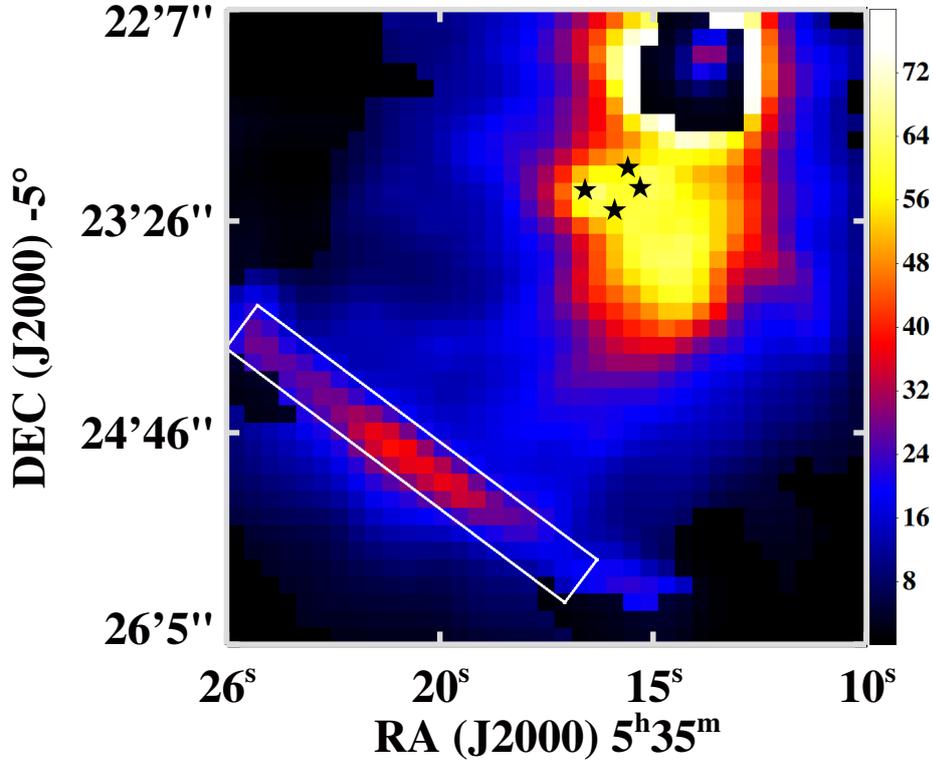

Fig. 9.— Total infrared luminosity from the blackbody fits, we extracted the luminosity in the aperture marked in white.

### 4.3. Dust extinction

In this Section, we determine the dust opacity in the ionized gas and PDR from the observations. The results show that the UV and IR opacities are about an order of magnitude less than for the diffuse ISM. In Section 5.3 we look at the implications.

### 4.3.1. The IR extinction in the HII region

We can estimate the IR extinction properties of the gas in the ionized gas by comparing our fits with the radio emission measure from Dicker et al. (2009) and the electron density. From the warm dust maps, we have selected a region where the higher temperature and morphology indicates that it is associated with the ionized gas (white square in Figure 3). At this position, the measured 19.7 $\mu$m optical depths is $0.025 \pm 0.003$, while the emission measure – estimated from the published emission measure map of Dicker et al. (2009) is $7 \pm 0.5 \times 10^6$ cm$^{-6}$ pc. With the measured electron densities ($2500 \pm 500$ cm$^{-3}$), we arrive at a column density of $8 \pm 2 \times 10^{21}$ cm$^{-2}$. This then translates into a 19.7 $\mu$m dust opacities of $3.1 \pm 0.9 \times 10^{-24}$ cm$^2$/H-atom. This value is a factor of about 10 less than the measurements by Chiar & Tielens (2006) towards Cyg OB 2 and about a factor 4 less than in typical models for interstellar dust (1.3 to $1.6 \times 10^{-23}$ cm$^2$/H-atom; Draine 2003).

### 4.3.2. The UV extinction in the HII region

In the HII region, dust grains and hydrogen compete for the absorption of stellar radiation but Lyman $\alpha$ will be (locally) absorbed by the dust. As in our previous study (Salgado et al. 2012), we can estimate the UV extinction produced by dust in the ionized region. The total IR luminosity associated with the ionization bar is measured to be $2.6 \times 10^3$ L$_\odot$ by integrating our modified-blackbody fits in this part of the HII region. From the emission measure image presented by Dicker et al. (2009), we estimate an average $EM$ of $6 \times 10^6$ cm$^{-6}$ pc, which corresponds to a Lyman $\alpha$ luminosity of $1.8 \times 10^3$ L$_\odot$. Hence, the dust heating due to absorption of star light is small in the ionized bar (800 L$_\odot$). From the geometry derived in Section 3.2, we estimate that the fraction of the stellar luminosity available corresponds to $A_{abs}/4\pi d^2$. We use $A_{abs} = 0.09$ pc$^2$ which goes with an actual distance of 0.45 pc and a luminosity of $2.7 \times 10^5$ L$_\odot$ to arrive at the dust UV optical depth, $\tau_{UV}$, of $8 \times 10^{-2}$. With the width of this region (0.05 pc) and a density of $2.5 \times 10^3$ cm$^{-3}$ (Wen & O'Dell 1995), this results in a UV dust opacity of $2 \times 10^{-22}$ cm$^2$/H-atom. These values can be compared to the calculated UV opacity of dust in the diffuse ISM, which – for an $R_V$ of 5.5 – ranges from 1 to $1.5 \times 10^{-21}$ cm$^2$/H-atom over the wavelength range of $2200 - 912$ Å (Weingartner & Draine 2001; Draine 2003). Similar values are actually obtained for sight-lines in the diffuse ISM characterized by $R_V = 3.1$. Hence, the UV opacity of dust in the Orion HII region is a factor of $\simeq 5$ less than in the diffuse ISM.

### 4.3.3. IR Extinction in the PDR

The 19.7 $\mu$m optical depth of the warm dust in the Orion Bar PDR is measured to be 0.2 (Figure 8). Taking the H-column density derived from the CII observations ($7 \times 10^{22}$ cm$^2$; Ossenkopf et al. (2013); see Section 4.1), we arrive at a 19.7 $\mu$m extinction per H-atom of $2.9 \times 10^{-24}$ cm$^2$/H-atom. These values are a factor 10 less than the values measured towards Cyg OB2 ($3.7 \times 10^{-23}$

cm$^2$/H-atom Chiar & Tielens 2006) and a factor 5 less than the models from Draine (2003) (1.3 to 1.6 × 10$^{-23}$ cm$^2$/H-atom, depending on the value for $R_V$).

### 4.3.4. The Penetration of UV Photons in the PDR

Since PAHs are excited by FUV photons, the decrease in PAH emission traces the dust grain extinction through the PDR. The emission of PAHs at a distance $d$ is given by:

$$F_{PAH}(d) = F_{PAH}(peak) \times e^{-\tau_{UV}(radial)}, \tag{1}$$

with $\tau_{UV}(radial) = N_d^T \times \sigma_{UV}$ the radial optical depth. The PAH emission decreases by a factor of 8 over a length of 30″ (0.06 pc) from the peak emission. With a density of 10$^5$ cm$^{-3}$, this translates into a UV opacity of 1.6 × 10$^{-22}$ cm$^2$/H-atom. Hence, as for the dust in the ionized gas, the dust in the PDR is characterized by a UV opacity which is about a factor 10 less than for dust in the diffuse ISM.

## 5. Discussion

### 5.1. The Dust Temperature and Lyα Heating

It has been long suspected that trapped Lyα photons can be a major source of dust heating in HII regions (e.g. Wright 1973; Garay et al. 1993; Smith et al. 1999). As the discussion in Section 4.3.2 demonstrates, Lyα photons are very important for the heating of dust in the Orion Nebula. Here, we will compare the derived temperatures of the warm dust component. The total heating of the dust depends on the distance to the central star, the radial extinction of UV photons towards the star and the number of trapped Lyα photons. These two sources of heating lead to a different distance dependence for the temperature. Because of resonant scattering, Lyα photons are absorbed on the spot and the contribution to the heating rate is,

$$\Gamma_{Ly\alpha} = n_e n_H f \alpha_H^{(2)} h\nu_\alpha / n_d, \tag{2}$$

with $n_e$ the electron density, $n_H$ the hydrogen density, $n_d$ the dust density, $h\nu_\alpha$ the energy of the Lyα photon (10.2 eV), $f = 0.7$ is fraction of recombination to levels higher than 2 that lead to emission of a Lyα photon, and $\alpha_H^{(2)}$ is the recombination coefficient of Hydrogen (case B). In addition, there is the direct stellar radiation,

$$\Gamma_d(a, r) = \pi a^2 \frac{L_\star(UV)}{4\pi r^2} e^{-\tau_{radial}} Q_{abs}(UV) + \Gamma_{Ly\alpha} \tag{3}$$

where $\tau_{radial}$ is the UV optical depth from the star to a point $r$ and $a$ is the radius of a spherical grain. Note that the stellar UV luminosity in terms of $G_0$ is $L_\star(UV)e^{-\tau_{radial}}/4\pi R_s^2$. The dust temperature can be obtained by equating the heating with the wavelength integrated emission of a dust grain. The emission can be approximated well by:

$$\Gamma_{em} = 4\pi a^2 Q_0 \sigma T_d^6, \tag{4}$$

where $\sigma$ is the Stefan-Boltzmann constant and for the constant $Q_0$ we adopt the properties of silicates $[Q_0 = 1.25 \times 10^{-5}(a/0.1\,\mu m)^{0.06}$, Tielens 2005]. The dust temperature is then given by:

$$T_d^6 = \frac{Q_{abs}(\text{UV})G_0}{4Q_0\sigma}\left[\frac{3h\nu_\alpha f}{\tau_d \overline{h\nu}} + \left(\frac{R_s}{r}\right)^2\right], \tag{5}$$

where $\overline{h\nu}$ is the average photon energy per ionizing photon ($L_\star/N_{ion} \simeq 49$ eV, for the parameters of $\theta^1$ Ori C), $R_s = 0.23$ pc is the Strömgren radius and $\tau_d = 0.08$ is the dust optical depth. We have adopted $G_0 = 2.6 \times 10^4$ (Section 4.1), $Q_{abs}(\text{UV}) \approx 1$. We emphasized that all these parameters – except for the detailed absorption/emission properties of the dust – are directly derived from observations. Note that the calculated dust temperature depends on the ratio of the UV absorption to the IR emission cross section. While we have derived a much reduced dust opacities in Section 4.3, the ratio of the two is actually similar to what we have adopted here.

The results are compared to the observed temperature in Figure 10. The calculated temperatures show the expected characteristic of Lyman $\alpha$ heating; i.e., constant dust temperature throughout the ionized gas. Heating by the photons from the central star does not start to dominate the dust temperature until $r \lesssim 0.25\ R_s$. We note that if we exclude Lyman $\alpha$ heating, the geometric dilution of the radiation field leads to a noticeable decrease in the dust temperature throughout the region. This is not present in the data again illustrating the importance of Lyman $\alpha$ heating of the dust. The absence of an upturn in the observed temperature at small radii likely reflects the absence of dust close to $\theta^1$ Ori C as the stellar wind may have created a dust free cavity. Finally, in order to reproduce the observed temperatures, dust grains with sizes between $0.1 - 1\ \mu m$ are required. This value is somewhat large compared to the dust in the diffuse ISM. We discuss this further in Section 5.2.

## 5.2.   Properties of the Dust

We have determined opacities in the UV and IR – in units of cm$^2$/H-atom – both for dust in the ionization bar and in the PDR (c.f., Section 4.3). As emphasized before, the derived UV and IR opacities are reduced by a factor $5 - 10$ as compared to dust in the diffuse ISM. Recently, Planck Collaboration et al. (2014b) reported maps of the spectral index of the whole sky including the Orion Nebula, where they report a value of $\beta = 1.6$. Using the *Planck* value ($\beta = 1.6$), we obtain differences in the temperatures of dust for the cold and warm components of $\sim 7\%$ and $\sim 5\%$, respectively. The small difference in temperature further reduces the column densities and, thus the opacities, by a factor of $\sim 3$. We attribute the decreased opacity values to the effects of coagulation. Indeed, theoretical studies show that coagulation can greatly reduce the extinction per unit mass both in the UV and the IR, if the grains can grow to very large sizes (Ossenkopf & Henning 1994; Ormel et al. 2011). A similar result was obtained by Lombardi et al. (2014) analyzing the near infrared extinction and sub-millimeter *Planck* maps for the Orion A and B clouds. Detailed models show that a factor of 10 reduction in the opacity is reached after a time, $t \simeq 100\,\tau_0$ where $\tau_0$ is the collision timescale at the onset of coagulation,

$$\tau_0 = (n_d \Delta v \sigma_d)^{-1} \simeq 8.5 \times 10^4 \text{ yr } \left(\frac{a_0}{0.1\,\mu\text{m}}\right)^{1/2} \left(\frac{n}{10^5\,\text{cm}^{-3}}\right)^{-3/4} \left(\frac{T}{10\,\text{K}}\right)^{-1/4}, \qquad (6)$$

with $n_d$, $\Delta v$ and $\sigma_d$ are the number density, relative velocity and cross section of the grains (Ormel et al. 2009). In the right-hand-side of this equation, an initially normal dust-to-gas ratio was adopted and $a_0$, $n$, and $T$ are the size of the monomer, gas density, and temperature, respectively. At this point ($t \simeq 100\,\tau_0$), the calculated porosity of the aggregates is $\simeq 0.3$ and the aggregate size is $\sim 10$ $\mu$m. From this timescale, we infer that the large grain size must reflect coagulation before the formation of $\theta^1$ Ori C, the HII region, and the PDR; i.e., during the cold dark cloud phase of Orion A. During this phase, coagulation is much assisted by the presence of ice mantles. However, when the dust temperature reaches some 75 K, the ice will sublimate on a timescale of $10^4$ yr, but we surmise that ice sublimation associated with such gentle thermal processing will not affect the structure of the dust aggregates. Once the dust leaves the shielded environment of the PDR and enters the HII region, further processing may occur. Sputtering will have little effect at $T = 10^4$ K but grain-grain collisions, even at $\simeq 5$ m s$^{-1}$, may lead to fragmentation (Wada et al. 2013; Krijt et al. 2015). Our results suggest, though, that this processing takes place on a timescale long compared to the evolution of the HII region. Large dust grains seem to be a common characteristic of dense cloud cores as evidenced from near-infrared scattered light images (e.g.: Pagani et al. 2010; Steinacker et al. 2010). Likewise, analysis of the IR emission from the HII region, IC 434, also inferred the presence of large dust aggregates in the champagne flow from the dark cloud, L1630 Ochsendorf et al. (2014a).

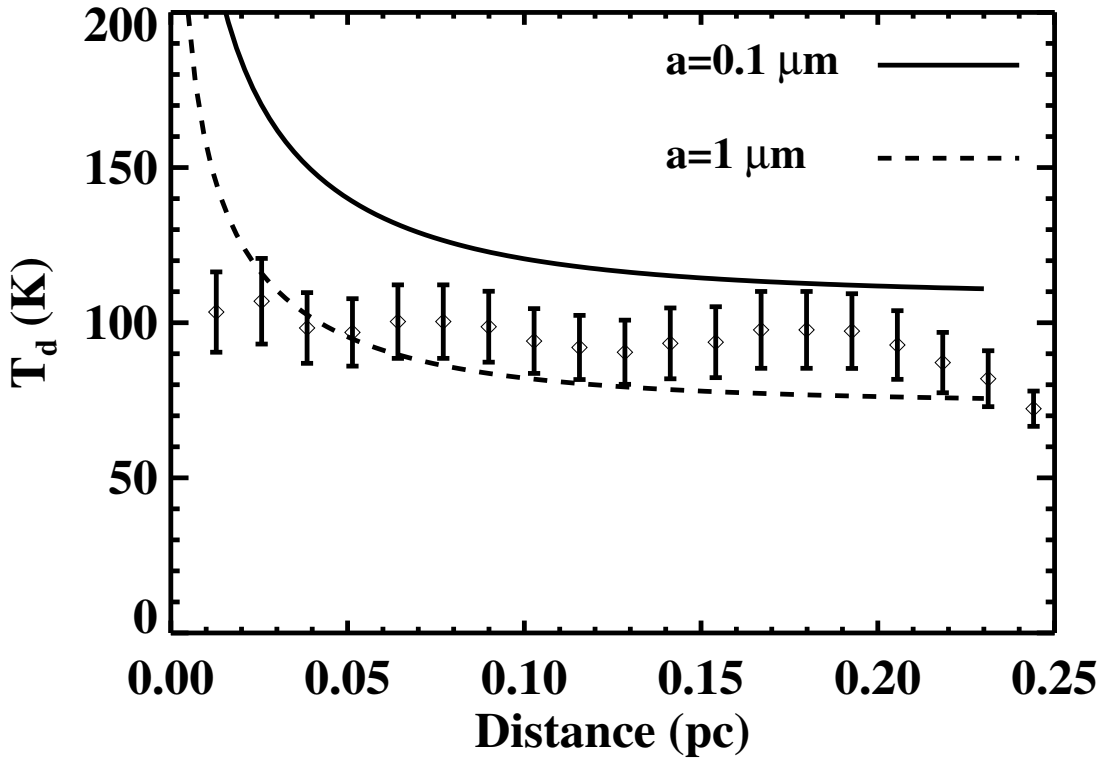

Fig. 10.— Temperature as a function of distance for different dust grain size: 0.1 $\mu$m (solid line) and 1 $\mu$m (dashed line). Symbols represent derived dust temperatures of the warm component.

## 5.3. Dust in HII Regions

There has been a long history of infrared studies of dust in HII regions (e.g.: Ney et al. 1973; Harper 1974). Here, we will place our analysis of the Orion HII region in context with a recent study of the W3(A) HII region and the results from the GLIMPSE survey.

In a previous study, Salgado et al. (2012) analyzed the dust emission in W3(A) using the same mid-infrared FORCAST filters. Like Orion, W3(A) is a young, compact HII region and is powered by a similar ionizing star. The ionized gas emission in the Orion nebula, as traced by H$\alpha$, is high close to the Trapezium stars then it slowly decreases up to 0.18 pc where it sharply increases, while at larger distances it sharply decreases again (Figure 4). Analysis of the optical emission morphology and spectra of the HII region reveals that the ionized gas is distributed in a thin shell (0.02 pc) bounded at the outside by the ionization front (Wen & O'Dell 1995). The ionized gas (as traced by free-free emission at 2 cm) in W3(A) has a similar structure, but the ionized gas shell is much thicker ($\simeq 0.1$ pc).

When comparing the dust emission of Orion with W3(A) we also see some similarities in the spatial distribution. In Orion, the 19 $\mu$m emission shows a component coinciding with the Orion Bar and one that is associated with dust in the ionized gas (Figure 3; Section 3.2). The other IR dust/PAH tracers only peak up in the Orion Bar PDR. For W3(A), the 19.7 $\mu$m emission traces dust well mixed in with the ionized gas shell. As for Orion, at other wavelengths, the dust (and PAH) emission clearly peak in the PDR surrounding the ionized shell. Hence, for both regions, the 19.7 $\mu$m flux originates from dust mixed in with the ionized gas. We note that unlike Orion, the 19.7 $\mu$m map of W3(A) does not show a secondary peak at the PDR (Salgado et al. 2012). In the PDR the dust is exclusively heated by stellar photons and this difference in 19.7 $\mu$m morphology merely reflects the smaller scale size and concomitantly higher dust temperature for Orion as compared to W3(A).

The heating of the dust is however different for these two HII regions. In Orion, resonantly scattered Lyman $\alpha$ photons contribute some 2/3 of the dust heating in the ionized gas; the remainder is due to absorption of stellar FUV photons. In contrast, for W3(A), dust heating is dominated by stellar photons with only a minor contribution from Lyman $\alpha$. Our analysis shows that this reflects the very low optical depth of dust in the ionized gas in Orion ($\tau_{UV} \simeq 0.08$). For W3(A), the UV optical depth was measured to be $\simeq 1$ and consequently stellar heating is much more important (cf., Equation 2). To phrase it differently, irrespective of the dust properties, all the Lyman $\alpha$ energy will eventually wind up in the dust. In contrast, the absorption of stellar photons depends strongly on the dust opacity per H-atom and, for these two regions, the derived UV dust opacities (per H-atom) differ by almost a factor 100.

This difference in dust opacity per H-atom must reflect a difference in evolution between these two regions. Both stars formed in dense cores where coagulation had likely reached a steady state. Both HII regions are also very young: W3(A) is slightly larger but Orion is already optically visible. Morphologically speaking, for W3(A), the shell structure likely betrays the importance of radiation pressure on the ionized gas and the dust, although we recognize that theoretical models predict a much shallower rise in the shell density than observed (Draine 2011). In contrast, for Orion, the HII region has broken out of the molecular cloud and has created a champagne flow (Güdel et al. 2008) and the dust is dragged along (Ochsendorf et al. 2014a). The difference in dust properties might therefore originate in or relate to the champagne flow phase in the evolution of compact HII regions that W3(A) has not yet entered in. Possibly it therefore reflects a size-sorting of the dust due to radiation pressure during the blister phase (Ochsendorf & Tielens 2015).

The Spitzer/GLIMPSE survey has revealed some 6000 bubble HII regions in the Milky Way. These are invariably characterized by 24 $\mu$m emission associated with dust inside the bubble while emission at 8 $\mu$m and at far-infrared wavelength traces the PDR surrounding the HII region (Churchwell et al. 2006, 2007; Deharveng et al. 2010). The morphology of the compact HII regions, Orion and W3(A), is very similar where dust in the ionized gas is visible at 19.7 $\mu$m while the PDR is traced in the PAH and colder dust emission. As the spectral energy distribution of the dust in the ionized gas peaks around 20 $\mu$m in all cases, the dust must have very similar temperatures, suggesting that Lyman $\alpha$ heating may dominate dust heating in HII regions in general. Alternatively, the 24 $\mu$m emission in these bubbles may trace a dust wave where the radiation pressure of the stellar light stops onrushing dust in a champagne flow and forces it in an arc-like structure around the star (Ochsendorf et al. 2014a,b). As the stellar properties are very similar, the dust wave may form at very similar distances, resulting in very similar temperatures.

## 5.4. Cooling Lines and Photoelectric Heating Efficiency

In dense PDRs, gas cooling is dominated by emission of forbidden transition lines of [O I] at 63 $\mu$m and [C II] at 157 $\mu$m (Hollenbach & Tielens 1999). Observations of the gas cooling lines in the Orion Bar have been discussed extensively by Bernard-Salas et al. (2012). Here we reanalyzed the data, including the mid-infrared contribution to the SED and the total IR emission of this region. In PDRs, the dominant heating process is due to the photoelectric effect in PAHs and small grains (de Jong et al. 1980; Bakes & Tielens 1994). The efficiency of the process can be measured directly from the ratio between the main gas cooling lines and the total energy absorbed by PAHs and dust grains.

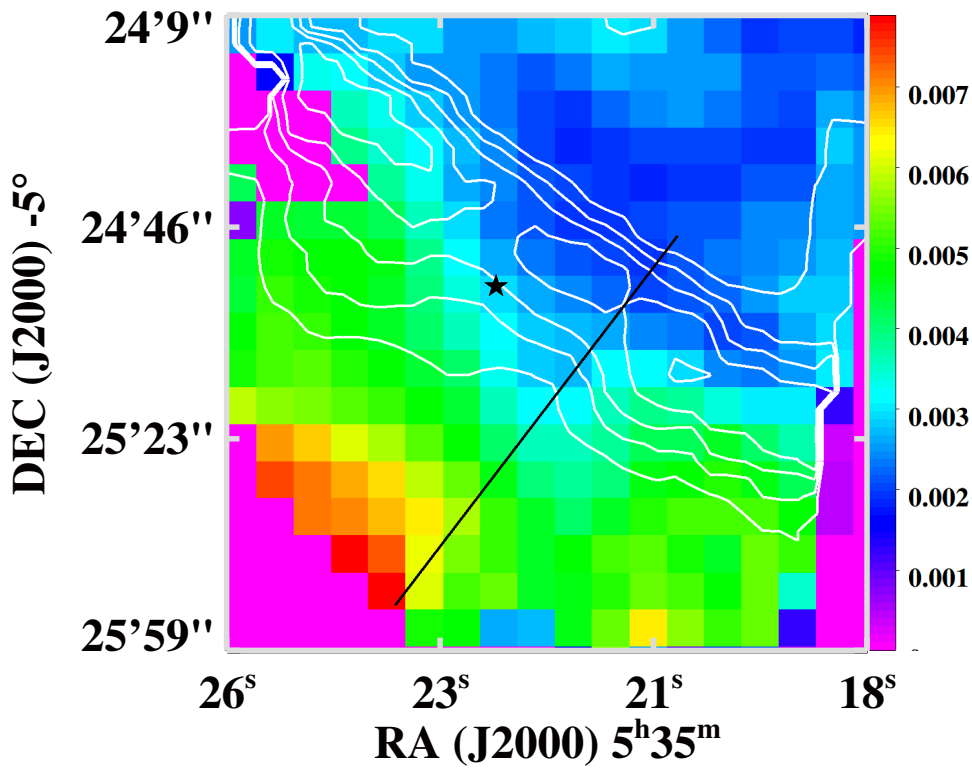

Fig. 11.— Photoelectric heating efficiency map, the star marks the position of $\theta^2$ Ori A and the line corresponds to the crosscut for regions in Figure 12. The crosscut extends from the surface of the PDR to deep in the molecular cloud, with a position angle of 143° measured counter-clockwise from north. Overlaid in white is the contour map of [O I], the main cooling line. The photoelectric heating efficiency increases with distance from the trapezium stars.

We used the PACS [O I] at 63 and 145 $\mu$m, and the [C II] at 157 $\mu$m lines to determine the photoelectric heating efficiency, i.e. the ratio between [O I]$_{63}$+[C II]$_{157}$+[O I]$_{145}$ and the total IR emission derived from the blackbody fits (Section 3.3). Typical flux values for the lines are $7\times10^{-2}$ and $6\times10^{-3}$ erg s$^{-1}$ cm$^{-2}$ sr$^{-1}$ for the [O I] at 63 $\mu$m and 145 $\mu$m, respectively. For the [C II] line, a typical value is $8\times10^{-3}$ erg s$^{-1}$ cm$^{-2}$ sr$^{-1}$. In the Orion Bar, we find an efficiency of $2\times10^{-3}$ which increases further into the PDR and can reach values as high as $7\times10^{-3}$ (see Figure 11).

The values found here are similar to those derived from Table 3 in Herrmann et al. (1997) and for M17 (Meixner et al. 1992). In this analysis, we assume that the far-infrared dust continuum emission is a good measure for the UV flux locally absorbed. That is a good assumption at the surface of the PDR, where FUV photons dominate. However, deeper in, the stellar UV flux is attenuated and absorption of mid-infrared photons produced by warm dust in the surface layers of the PDR become an important dust heating source. The effect of this is that the actual increase of the heating efficiency (defined as gas heating over absorbed FUV flux, Tielens & Hollenbach 1985) increases even more than the factor 2 indicated above. This increase in the heating efficiency with depth in to the PDR is a standard feature of photo-electric heating models. As the UV field decreases, the PAHs and very small dust grains become less positively charged and the heating efficiency increases (Bakes & Tielens 1994).

In Figure 12, we show the photoelectric efficiency as a function of the ionization parameter, $\gamma = G_0 \sqrt{T}/n_e$ along a line perpendicular to the Orion Bar surface, which avoids – as best as possible – the clumps present in the region. In this analysis, we assume that the region is homogeneous in temperature ($T = 500$ Parmar et al. 1991; Allers et al. 2005) and an electron density ($n_e = X_e n_H$) with an electron abundance equal to the gas phase carbon abundance ($X_e = 1.5 \times 10^{-4}$; Cardelli et al. 1996) and $n_H = 10^5$ cm$^{-3}$ (Simon et al. 1997). The variation in the ionization parameter is then largely due to variation in the penetrating UV field, which we approximate by the incident radiation field ($G_0 = 2.6 \times 10^4$ in Habing units, Section 4.1) extincted by $\tau_{UV} = 8(d/30'')$ (Section 4.3.4). The figure illustrates well that as the ionization parameter increases, the heating efficiency decreases. A similar trend was also noted by Okada et al. (2013) in their analysis of a sample of 6 PDRs. However, as their data set is very inhomogeneous and the ionization parameters are difficult to determine in an absolute sense, the trend is less clear. In this study, we focus on a single region where variations in the ionization parameter are driven by the drop of the radiation field and a clear trend is revealed. We do note though that the absolute scale of the ionization parameter does depend on the adopted temperature and density. It is reasonable to assume that the region is in pressure equilibrium and hence the ionization parameter would scale as $G_0/\sqrt{T}$. We have indicated points beyond the peak of the H$_2$ emission where this variation in temperature would become relevant as red in Figure 12. The ionization parameter at these locations might have been underestimated by up to a factor $\simeq 2.5$ (i.e., $\sqrt{T_{H_2}/T_{CO}} = \sqrt{500/75}$).

We recognize the general increase in the heating efficiency as predicted by model studies (Bakes & Tielens 1994; Weingartner & Draine 2001). However, the observed heating efficiency differs from that expected from models. The theoretical models were developed for the diffuse ISM where observed heating efficiencies are higher (Wolfire et al. 1995; Hollenbach & Tielens 1999). This difference in heating efficiency may just reflect a difference in dust properties; i.e., a lower abundance of PAHs and very small grains in the Orion Bar PDR (Section 5.2).

The photoelectric heating efficiency is expected to be linked to the degree of ionization of the PAHs and very small grains. For PAHs, the degree of ionization can be traced by the ratio between the 7.7 $\mu$m and the 11.3 $\mu$m features as the 7.7 $\mu$m feature is emitted by ionized PAH and the 11.3 $\mu$m by neutral PAHs (DeFrees et al. 1993; Langhoff 1996; Allamandola et al. 1999). Following Galliano et al. (2008), we used the ISOCAM-CVF data to get values for the ionized to neutral ratio of PAHs (Figure 12). As expected, there is a clear increase in the fraction of PAH cations from the $H_2$ peak to the front of the PDR, in good agreement with the trend found by Galliano et al. (2008). However, we notice that the PAH ratio reaches a plateau starting at the $H_2$ peak until the end of the dense bar (which we associate with the swept up shell). Perhaps, this reflects the influence of anions deep in the PDR.

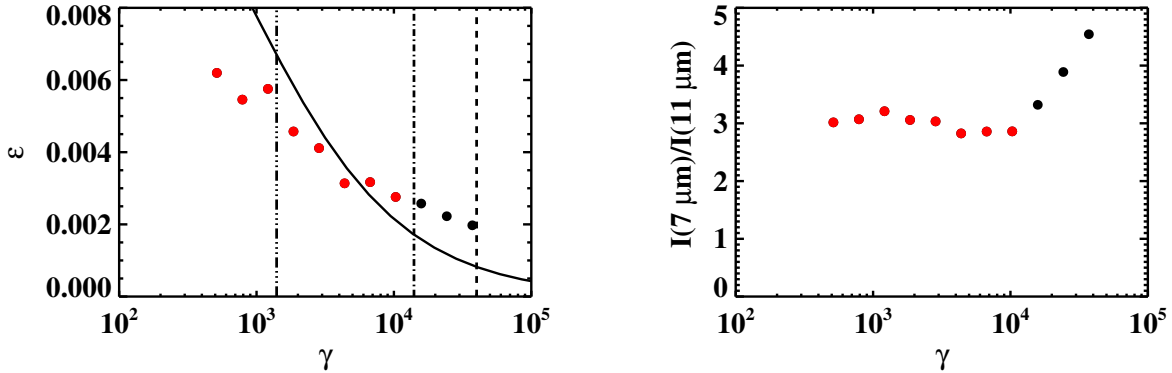

Fig. 12.— Photoelectric heating efficiency as a function of ionization parameter ($G_0\sqrt{T}/n_e$) along the cross cut through the Orion Bar (Figure 11). The dashed line marks the position of the ionization front, the dot-dashed line the position of $H_2$ peak and the dot-dot-dashed the end of swept up gas; the solid line is the model for photoelectric heating from Bakes & Tielens (1994) scaled sown by a factor of 3. $G_0$ was estimated from the incident radiation field attenuated in the Bar. The temperature ($T = 500$ K) and electron density ($n_e = 15$ cm$^{-3}$) were assumed to be constant (see text for details). Black dots refer to warm ($T \simeq 500$ K) gas traced by the pure rotational lines of $H_2$. The red dots are cooler molecular gas ($T \simeq 50 - 100$ K) traced by CO.

As noted in Section 3.2, both the cooling line and the dust emission display a clumpy distribution. We also see variations of the photoelectric heating efficiency parallel to the Orion Bar surface. However, the clumpy structure apparent in the cooling line maps is not seen in the photoelectric heating map (Figure 11). It seems that the increased density associated with clumps does not translate in an increased photoelectric heating efficiency. These observations illustrate that further detailed studies of the cooling lines and dust emission coupled with density probes (such as molecular tracers) are needed to probe the relationship between gas cooling and the PAH/dust components responsible for the heating.

## 6. Summary and Conclusions

We analyzed the infrared emission in the Orion star forming region in six bands observed with the FORCAST instrument on board of SOFIA. We complemented the dataset with public published photometry and spectroscopy ranging from optical to far-infrared wavelengths. The FORCAST images show a complex structure and we recognize emission arising in the Orion Bar PDR, the BN/KL object and the Ney-Allen nebula. Our analysis shows that the *Herschel*/PACS far-infrared emission is associated with the cold molecular cloud. The mid-infrared emission is not spatially correlated with the far-infrared emission and must be produced in the PDR and, partially, in the HII region.

We constructed SEDs after convolving the FORCAST images to the PACS 160 $\mu$m resolution. By fitting the SEDs with two modified blackbodies, we derived dust temperatures and column densities for dust associated with the molecular cloud and the PDR/HII region. Most of the dust mass is in the cold dust component which is associated with the background cold and dense molecular cloud. The luminosity, on the other hand, is dominated by the warm dust component associated with the PDR/HII region. We obtained the total infrared luminosity by integrating our blackbody fits over wavelength between 1 and 1000 $\mu$m. The results of the fits were used to obtain the geometry of the Orion Bar PDR. We measured a depth along the line of sight of 0.28 pc and an inclination angle of 4°.

We have determined that the Orion Bar extends from the ionization front located at 0.23 pc from $\theta^1$ Ori C up to the shock front located at about 0.1 pc from the ionization front. The amount of mass swept by the expansion of the HII region is $6 \times 10^3$ M$_\odot$ and we estimate an age of $\simeq 10^5$ yr for the HII region and $\theta^1$ Ori C. This age is much less than that derived by stellar evolutionary models (2.5 Myr, Simón-Díaz et al. 2006).

By comparing the 19.7 $\mu$m image with ionized gas radio emission we have determined the opacity (extinction per hydrogen atom) in the HII region to be a factor of 5 lower than models for the diffuse interstellar medium. In addition, we have computed the UV opacity in the ionized gas by comparing the infrared emission and radio estimates. As for the infrared opacity, the UV opacity value is a factor of 5 lower than that of the diffuse ISM.

We measured infrared opacity in the PDR by comparing the optical depth and the density. The value obtained by this method is a factor of 5 less than models for the diffuse ISM. By measuring the penetration of UV photons in the PDR, as traced by the PAH emission, we derived an UV opacity a factor of 10 less than that of the ISM.

Our analysis shows that Ly$\alpha$ heating is important in Orion at distances $r > 0.25R_s$. The dust temperature profile can be explained by the presence of large grains (between 0.1-1 $\mu$m).

The lower opacities as compared to ISM values and the Ly$\alpha$ heating behavior depend on the size of the dust grains. The observed values for the opacities and dust temperature in the ionized gas can be explained by the presence of large grains. We attribute the increase in size of the grains to dust coagulation during the molecular cloud phase previous to the birth of the ionizing stars.

We measure the photoelectric heating in the Orion Bar from the observed fine-structure cooling lines and the IR continuum emission. The observations reveal that the photo-electric efficiency decreases towards the surface of the PDR. Theoretical models attribute this to a charging-up of the PAHs and very small grains. We estimate values for the ionization parameter as a function of depth in the Orion Bar PDR. The derived heating efficiency as a function of ionization parameter agrees qualitatively well with the models but is less by about a factor 5. As the models were derived for the diffuse ISM, this quantitative difference may merely reflect a difference in PAH abundance.

We thank the FORCAST engineering team: George Gull, Justin Schoenwald, and Chuck Henderson, and the USRA Science and Mission Ops teams, and the entire SOFIA staff. F. S. would like to thank Prof. Alain Abergel and Prof. Malcolm Walmsley for useful comments and discussions during the preparation of this manuscript. This work is based on observations made with the NASA/DLR Stratospheric Observatory for Infrared Astronomy (SOFIA). SOFIA science mission operations are conducted jointly by the Universities Space Research Association, Inc. (USRA), under NASA contract NAS2-97001, and the Deutsches SOFIA Institut (DSI) under DLR contract 50 OK 0901. Financial support for FORCAST was provided by NASA through award 8500-98-014 issued by USRA. Studies of interstellar PAHs and dust at Leiden Observatory are supported through advanced-ERC grant 246976 from the European Research Council, through the Spinoza premie of NWO, and through the Dutch Astrochemistry Network funded by the Dutch Science Organization, NWO. This research has made use of the SIMBAD database, operated at CDS, Strasbourg, France. This research has made use of the NASA/ IPAC Infrared Science Archive, which is operated by the Jet Propulsion Laboratory, California Institute of Technology, under contract with the National Aeronautics and Space Administration.